\documentclass[aps,prd,onecolumn,superscriptaddress,nofootinbib]{revtex4}
\usepackage{epsfig,amsfonts,amsthm}
\usepackage{amsmath,amssymb}
\usepackage[usenames,dvipsnames]{xcolor} % p usar nomes de cores
\usepackage{hyperref}
\hypersetup{
   colorlinks=true,       % false: boxed links; true: colored links
    linkcolor=Blue,          % color of internal links
    citecolor=Plum,        % color of links to bibliography
    filecolor=magenta,      % color of file links
    urlcolor=YellowOrange           % color of external links
}
\usepackage[utf8]{inputenc}

\newcommand{\be}{\begin{equation}}
\newcommand{\ee}{\end{equation}}
\newcommand{\bea}{\begin{eqnarray}}
\newcommand{\eea}{\end{eqnarray}}

\renewcommand{\Im}{\mathrm{Im }}

\newcommand{\Tr}{\mathrm{Tr}}

\newcommand{\RR}{\mathbb{R}}

\providecommand{\id}{{\boldsymbol{1}}}

\definecolor{darkgreen}{rgb}{0.0, 0.6, 0.2}

\def\lsim{\mathrel{\rlap{\lower4pt\hbox{\hskip1pt$\sim$}}
    \raise1pt\hbox{$<$}}}         %less than or approx. symbol
\def\gsim{\mathrel{\rlap{\lower4pt\hbox{\hskip1pt$\sim$}}
    \raise1pt\hbox{$>$}}}         %greater than or approx. symbol

\providecommand{\eq}[1]{\begin{equation} #1 \end{equation}}
\providecommand{\aver}[1]{\langle #1 \rangle}
\providecommand{\eqali}[1]{\begin{equation}\begin{aligned} #1
    \end{aligned}\end{equation}}

\DeclareMathOperator{\aut}{Aut}

\DeclareMathOperator{\inn}{Inn}
\DeclareMathOperator{\adj}{adj}
\begin{document}
\title{
{\normalsize \hfill CFTP/19-028} \\*[7mm]
Discriminating between $CP$ and family transformations in the bilinear space of NHDM}

\author{Igor~P.~Ivanov}\thanks{E-mail: igor.ivanov@tecnico.ulisboa.pt}
\affiliation{CFTP, Instituto Superior T\'{e}cnico, Universidade de Lisboa,
Avenida Rovisco Pais 1, 1049 Lisboa, Portugal}
\author{Celso~C.~Nishi}\thanks{E-mail: celso.nishi@ufabc.edu.br}
\affiliation{Centro de Matem\'atica, Computa\c c\~ao e Cogni\c c\~ao,
Universidade Federal do ABC - UFABC,
09.210-170, Santo Andr\'e, SP, Brazil}

\begin{abstract}
The scalar potential of the $N$-Higgs-doublet model (NHDM) is best analyzed 
not in the space of $N$ complex doublets $\phi_a$ but in the $N^2$-dimensional space
of real-valued bilinears constructed of $\phi_a^\dagger \phi_b$.
In particular, many insights have been gained into $CP$ violation in the 2HDM and 3HDM 
by studying how generalized $CP$ transformations (GCPs) act in this bilinear space.
These insights relied on the fact that GCPs, which involved an odd number of mirror reflection, 
could be clearly distinguished from Higgs family transformations by the sign of the determinant
of the transformation matrix.
It was recently pointed out that this criterion fails starting from 4HDM,
where the reflection/rotation dichotomy does not exist anymore.
In this paper, we restore intuition by finding a different quantity which faithfully discriminates between
GCPs and Higgs family transformations in the bilinear space for any number of Higgs doublets.
We also establish the necessary and sufficient conditions for an orthogonal transformation
in the bilinear space to represent a viable transformation back in the space of $N$ doublets,
which is helpful if one prefers to build an NHDM directly in the bilinear space.
\end{abstract}

\maketitle
%\enlargethispage{1cm}

\section{Introduction}

$N$-Higgs-doublet models (NHDM) is a popular conservative framework 
of building models beyond the Standard Model (bSM).
Driven by the idea that the Higgs fields can also come in generations, just like fundamental fermions,
and by the numerous opportunities to impose new global symmetries,
such models allow one to resolve or alleviate many of the problems faced by the SM.
Sparked by the original proposals of the 2HDM by T.~D.~Lee in 1973 \cite{Lee:1973iz} 
and the 3HDM by S.~Weinberg in 1976 \cite{Weinberg:1976hu},
the field has evolved into a vibrant bSM playground, see for example book \cite{book} 
and recent reviews \cite{Branco:2011iw,Ivanov:2017dad}.

The rich phenomenology of NHDMs comes at the price of having to face technical challenges.
The scalar potential of the NHDM involves very many free parameters,
which is further complicated by the notorious issue of basis dependence.
Namely, when building models with several fields having identical quantum numbers such as the NHDM, 
one is allowed to perform an arbitrary unitary transformation mixing these fields
without changing the physical content of the model.
Although the Lagrangian and intermediate calculations may look vastly different in different bases, 
the phenomenological consequences must be the same.

Therefore, in order to efficiently explore all the phenomenological situations
offered by $N$ Higgs doublets, one must be able to tell which potentials 
are linked by a mere basis change and which represent truly distinct models. 
In short, one needs tools which allow one to explore the NHDMs in a basis independent way.

A particularly intriguing topic in the NHDM physics is the issue of $CP$ violation 
coming from the scalar sector \cite{book}.
A tricky issue here is that the $CP$ transformation itself can be defined in a variety of ways.
The standard definition in the space of complex scalar fields $\phi_a$, $a = 1, \dots, N$, is
$\phi_a(\vec{r}, t) \mapsto \phi_a^*(-\vec{r}, t)$.
However, this definition is basis dependent: 
the same physical consequences will follow from imposing
\begin{equation}
\phi_a(\vec{r}, t) \mapsto X_{ab} \phi_b^*(-\vec{r}, t)\,,\quad X \in U(N)\,,\label{GCP}
\end{equation}
provided the $CP$ transformation is of order 2 (denoted CP2), 
that is, applying it twice produces the identity transformation.
This extra condition restricts the matrix $X$ by the relation $XX^* = \id_N$.
However, one can also impose 
\eqref{GCP} without requiring it to be of order 2
\cite{Neufeld:1987wa,Ecker:1987qp,Grimus:1995zi}.
If, as a result of imposing this higher-order $CP$ symmetry, one does not acquire an accidental CP2, 
one obtains a new $CP$-conserving model which is truly different from the one based on CP2.
Examples of NHDMs have been constructed based on the $CP$ symmetry of order 4 (CP4) \cite{Ivanov:2015mwl}
and of higher orders \cite{Ivanov:2018qni}.

The vast experience gained in the 2HDM shows that it is much more convenient to discuss its $CP$ properties
not in the space of Higgs doublets $\phi_a$, $a=1,2$, but in the $1+3$-dimensional space of  
their gauge-invariant bilinear combinations $\phi_a^\dagger \phi_b$ grouped into the following real-valued variables:
\begin{equation}
r_0 = \phi^\dagger_a \phi_a\,,\quad r_i = \phi^\dagger_a (\sigma^i)_{ab} \phi_b\,,\quad i=1, 2, 3\,,
\label{bilinears-2HDM}
\end{equation}
where $\sigma^i$ are the familiar Pauli matrices.
The map (\ref{bilinears-2HDM}) from doublets $\phi_a$ to the $(r_0,r_i)$ covers not the entire $1+3$-dimensional space 
but only the forward cone defined by the inequalities 
\begin{equation}
r_0 \ge 0\,, \quad r_0^2 - r_i^2 \ge 0\,.\label{cone}
\end{equation}
A pure family transformation $\phi_a \mapsto \phi'_a=U_{ab} \phi_b$ with $U \in U(2)$ leaves $r_0$ 
invariant and induces an $SO(3)$ rotation of the vector $r_i$: 
\begin{equation}
r_i \mapsto R_{ij} r_j\,, \quad \det R = +1\,.\label{R}
\end{equation}
The standard $CP$ transformation $\phi_a \mapsto \phi_a^*$ induces the following planar reflection:
\begin{equation}
r_i \mapsto C^{(s)}_{ij} r_j\,, \quad C^{(s)} = \mathrm{diag}(1, -1, 1)\,, \quad \det C^{(s)} = -1\,.\label{Cs}
\end{equation}
A GCP transformation \eqref{GCP} induces a rotary reflection, which can always be presented as a product
of a pure rotation and the standard $CP$ transformation: $C = R C^{(s)}$.
Clearly,
\begin{equation}
\det C = \det R \cdot \det C^{(s)} = -1\,.\label{detC}
\end{equation}
Since the map $SU(2) \to SO(3)$ is surjective, any $O(3)$ transformation $O$ in the bilinear space can be realized 
either as a family transformation, if $\det O = +1$, or a GCP, if $\det O = -1$.

The geometric picture emerging from this dichotomy was discussed at length in \cite{Ivanov:2005hg,Nishi:2006tg,Ferreira:2010hy,Ferreira:2010yh}.
It gave a more direct proof of the basis-invariant conditions for the explicit $CP$ conservation in the 2HDM scalar potential
which had been previously established in \cite{Branco:2005em,Davidson:2005cw,Gunion:2005ja} 
with the aid of extensive computer algebra scans.\footnote{Very recently, the same $CP$-odd basis invariants were recovered via a powerful method based 
on Hilbert series and plethystic logarithm \cite{Trautner:2018ipq}, superseding the need for the brute force computer algebra checks.}
A detailed analysis of various paths leading to explicitly $CP$ conserving 2HDMs was presented in \cite{Ferreira:2010hy,Ferreira:2010yh}.
 
In 3HDM, the necessary and sufficient basis-invariant conditions for existence of a CP2 symmetry were first derived
in \cite{Nishi:2006tg} and recently put in a wider context in \cite{Ivanov:2019kyh,deMedeirosVarzielas:2019rrp}.
The meticulously elaborate analysis of the order-2 condition of the very recent study \cite{Maniatis:2019yqb} reconfirmed 
that CP2 can always be brought by a basis change to the standard form.
In addition, three Higgs doublets allow one to implement a novel type of $CP$ symmetry of order 4 \cite{Ivanov:2011ae,Ivanov:2015mwl},
and a basis-independent algorithm for detection of CP4 in 3HDM was given in \cite{Ivanov:2018ime}.

All these results were obtained only with the bilinear space formalism, stressing its superior role in analyzing the scalar sector of NHDMs.
Despite its complexity in the 3HDM as compared to the 2HDM, one still observes the discriminating role of the determinant:
all Higgs family transformations are described in the $1+8$-dimensional bilinear space by pure rotations $R \in SO(8)$ with $\det R = +1$,
while all GCPs induce transformations $C$, which involve an odd number of reflections, so that $\det C = -1$.

However when moving to 4HDMs, a peculiarity was recently pointed out in \cite{Maniatis:2019yqb}.
It turns out that in this case, both the family transformations and the GCPs induce transformations $O$ in the bilinear space with $\det O = +1$.
Thus, the intuition linking Higgs family transformations with pure rotations and GCPs with the orientation-flipping reflections
fails starting from 4HDM.

In this paper, we resolve this disturbing loss of intuition. 
Recalling that the bilinear space inherits the $SU(N)$ invariant tensors,
we replace the determinant with a different indicator which serves as a faithful discriminator 
between GCPs and family transformations in the bilinear space for any number of Higgs doublets.
As a by-product, we establish the necessary and sufficient conditions on an orthogonal transformation $O$
in the bilinear space of NHDM to represent a viable transformation back in the space of $N$ doublets,
either a basis change or a GCP. These conditions are especially helpful if one prefers to build an NHDM scalar sector 
directly in the bilinear space.

\section{GCPs vs. family transformations in the bilinear space}\label{section-bilinear}

\subsection{The problem}

Let us first remind the reader how the bilinear formalism is extended to 3HDM
\cite{Nishi:2006tg,Ivanov:2010ww,Maniatis:2014oza}. 
We define $1+8$ gauge-invariant bilinear combinations $(r_0, r_i)$:
\begin{equation}
r_0 = {1\over\sqrt{3}}\phi^{\dagger}_a\phi_a\,,\quad r_i = \phi^{\dagger}_a (t^i)_{ab}\phi_b\,,\quad 
i=1,\dots,8\,,\quad a=1,2,3\,.\label{bilinears}
\end{equation}
If $\lambda_i$ are the standard Gell-Mann matrices, then $t_i = \lambda_i/2$ are the generators of the $SU(3)$ algebra satisfying
$[t_i,t_j] = i f_{ijk} t_k$ and $\{t_i,t_j\} = \delta_{ij}\id_{3}/3 + d_{ijk} t_k$,
with the $SU(3)$ structure constants $f_{ijk}$ and the fully symmetric $SU(3)$ invariant tensor $d_{ijk}$.
These bilinears satisfy not only \eqref{cone} but also an additional constraint \cite{Ivanov:2010ww}:
\begin{equation}
d_{ijk}r_ir_jr_k + {1\over 2\sqrt{3}}r_0(r_0^2 - 3 r_i^2) = 0\,. \label{ddd}
\end{equation}
A basis change in the space of Higgs doublets $\phi_a \mapsto \phi'_a = U_{ab} \phi_b$ with $U \in SU(3)$ 
induces an $SO(8)$ rotation $R$ of the vector $r_i$, with $\det R = +1$.
With the usual notation for the Gell-Mann matrices,
the standard $CP$ transformation acts on $r_i$ by flipping components $2, 5, 7$:
\begin{equation}
r_i \mapsto C^{(s)}_{ij} r_j\,, \quad C^{(s)} = \mathrm{diag}(1, -1, 1, 1, -1, 1, -1, 1)\,,\label{Cs3}
\end{equation}
so that $\det C^{(s)} = -1$ still holds. 
By the same logic as before, any GCP transformation \eqref{GCP} can be written as a product of the standard $CP$ and a rotation,
which again leads to $\det C = -1$.
The geometric interpretation of GCP transformations becomes more involved \cite{Nishi:2006tg,Ivanov:2019kyh,Maniatis:2019yqb},
but the determinant still offers a faithful distinction between GCP transformations and pure basis changes.

The construction can be repeated for any number of Higgs doublets $N$.
The $N^2-1$-dimensional real-valued vector $r_i$ is constructed in a similar way based on traceless hermitean matrices $\lambda_i$,
\cite{Nishi:2006tg,Ivanov:2019kyh,Maniatis:2019yqb}.
Out of them, there are $N-1$ diagonal matrices and $N(N-1)/2$ pairs of the off-diagonal ones,
one real symmetric and one imaginary antisymmetric in each pair, 
for an explicit numbering scheme see, for example, \cite{Maniatis:2019yqb}.
The components of $r_i$ do not fill the entire space $\RR^{N^2-1}$ but lie in an algebraic manifold called the orbit space  
and defined, in addition to Eqs.~\eqref{cone} and \eqref{ddd}, by a series of 
algebraic equalities of orders up to $N$ with coefficients constructed from $SU(N)$-invariant tensors \cite{Ivanov:2010ww}.

All Higgs family transformations and GCPs of the NHDM are represented by those orthogonal transformations from $O(N^2-1)$
which leave this orbit space invariant.
The standard $CP$ transformation $\phi_a \mapsto \phi_a^*$ acts on the components of $r_i$ by $\pm 1$ factors,
with a $-1$ factor corresponding to each imaginary antisymmetric generator $\lambda_i$ which produces $r_i = \Im \phi_a^\dagger \phi_b$
for an suitable pair of doublets.
As a result, the determinant of the standard $CP$ transformation depends on the number $n_a$ of such antisymmetric generators\,\cite{Nishi:2006tg}:
\begin{equation}
\det C^{(s)} = (-1)^{n_a}\,, \quad n_a = \frac{N(N-1)}{2}.
\end{equation}
This determinant is $-1$ for $N=2, 3$ but changes to $+1$ for $N=4, 5$. 
This pattern ``two $-1$'s followed by two $+1$'s'' repeats itself as $N$ grows further \cite{Maniatis:2019yqb}.
Clearly, the same applies to all GCPs.

We are now ready to formulate the problem. Suppose we work in the bilinear space of NHDM and we are given an orthogonal transformation
$O \in O(N^2-1)$ that leaves the orbit space invariant. Is there a simple way to tell whether
it corresponds to a GCP or a Higgs family transformation?
In 2HDM and 3HDM, we developed the geometric intuition: all family transformations have $\det O = +1$,
all GCPs have $\det O = -1$. Now we see that this geometric intuition is lost already for 4HDM,
where {\em all} allowed transformations of Higgs doublets are represented 
by orientation-preserving $SO(15)$ rotations in the bilinear space. 
What is then the true faithful discriminator between GCPs and family transformations within the bilinear space?

\subsection{A resolution}

Let us first mention that working with traces of (powers of) matrices instead of the determinant 
does not help much.
Although $\Tr R$ is a basis-invariant quantity, traces of {\em different} basis changes are also different: 
$\Tr R_1 \not = \Tr R_2$. The same applies to traces of GCP transformations.
Moreover, since $\Tr AB \not = \Tr A\cdot \Tr B$, we cannot factor the trace of GCP into the traces of a rotation 
and the standard $CP$. Finally, it may happen that $\Tr R = \Tr C$, and one would need to compute traces
of their higher powers to distinguish the two.
In short, traces give too much detail of the transformation and are not suitable for a quick distinction 
between GCPs and basis-change.\footnote{We are not claiming that it is {\em impossible} to distinguish the two classes 
of transformations using traces of powers of matrices. It may be possible through computation 
of several such traces and evaluating a sophisticated function. We do not address this problem here 
because the quantity we show in the main text immediately solves the problem.}

Now, let us recall that the bilinear space is not the structureless $\RR^{N^2-1}$ but inherits the $SU(N)$ structure.
Therefore, if we are given an orthogonal transformation matrix $O$ which leaves the orbit space invariant,
not only are we allowed to evaluate traces or determinants 
but we can also contract it with $SU(N)$ invariant tensors $f_{ijk}$ or $d_{ijk}$.
For example, CP odd invariants are easily constructed with the use of $f_{ijk}$\,\cite{{Nishi:2006tg}}.
We propose the following invariant quantity as a discriminator between family transformations and GCPs:
\begin{equation}
J = \frac{1}{N(N^2-1)} f_{i'j'k'} f_{ijk} O_{i'i} O_{j'j} O_{k'k}\,.\label{O}
\end{equation}
If $O = R$ is a pure family transformation, then, due to the property of $f_{ijk}$ being invariant under $SU(N)$
basis changes $f_{i'j'k'} R_{i'i} R_{j'j} R_{k'k} = f_{ijk}$, we get
\begin{equation}
J_R = \frac{1}{N(N^2-1)} f_{ijk} f_{ijk} = 1\,.\label{O-R}
\end{equation}
Instead, if $O = C = R\cdot C^{(s)}$ is a GCP transformation, then 
\begin{equation}
J_C = \frac{1}{N(N^2-1)} f_{i'j'k'} f_{ijk} C^{(s)}_{i'i} C^{(s)}_{j'j} C^{(s)}_{k'k} = -1\,.\label{O-C}
\end{equation}
The last equality comes from the fact that the tensor $f_{ijk}$ is non-zero 
only when its indices $i, j, k$ involve an {\em odd} number of sign-flipping components.
This can be seen, for example, directly from the definition of the structure constants $f_{ijk}$,
which link three generators through the imaginary unit.
Therefore, each term in the implicit summation of \eqref{O-C} comes with 
$C^{(s)}_{i'i} C^{(s)}_{j'j} C^{(s)}_{k'k} = -1$ or $(-1)^3$.
These results hold for any $N$. 
Thus, $J$ defined in \eqref{O} serves as a faithful discriminator between GCPs and family transformations.

It is interesting to note that for $N=2$, the 3D orbit space recovers the full spherical symmetry.
As a result, all $O(3)$ transformations in the $r_i$ space can be induced either by a basis change or a GCP transformation.
On the other hand, the $SU(2)$ structure constants are $f_{ijk} = \epsilon_{ijk}$,
so that the invariant $J$ coincides with the well-known definition of the determinant.
This is why the intuition based on rotation/reflection worked in the 2HDM.

\subsection{Which transformations can be used in the bilinear space?}

Suppose one wants to build an NHDM and to implement a symmetry directly in the bilinear space.
Which orthogonal transformations $O$ can one use?
The quantity $J$ in Eq.~\eqref{O} already gives a strong constraint: 
only those $O$ which produce $J = \pm 1$ may be used as symmetries of the model.
If one wants to implement an $O$ with $\det O = +1$ but $|J| < 1$, 
then it cannot be induced by any Higgs family or GCP symmetry in the space of doublets.
Thus, $|J| = 1$ and $|\det O| = 1$ represent a necessary condition for the transformation $O$
to be a valid choice of a symmetry of the model. But is it sufficient?

The answer is no, and it can be illustrated with the following 5HDM example.
Consider the $C^{(s)}$ transformation in the 5HDM. 
In the bilinear space, which is 24-dimensional, it flips 10 directions, thus $\det C^{(s)} = +1$ but $J = -1$.
Now notice that the orthogonal transformation $O = - \id_{24}$ satisfies the same conditions:
$\det O = +1$ and $J = -1$. However, $O = - \id_{24}$ cannot be induced by any basis change or GCP back in the Higgs doublet space. 
The simplest way to see it is to observe
that the overall sign flip $r_i \to - r_i$ does not leave \eqref{ddd} invariant,
hence it does not preserve the orbit space.
In a similar way, one can consider the orthogonal transformation $O = - C^{(s)}$ in 5HDM, 
which again cannot be induced by any basis change or GCP but which satisfies $\det O = +1$ and $J = +1$,
the conditions one normally expects from regular basis changes.

We prove in the appendix that it is sufficient to complement the $|J|=1$ and $|\det O| = 1$ conditions
with the following extra requirement: $J_d = 1$, where
\eq{
	J_d\equiv \frac{N}{(N^2-1)(N^2-4)}d_{i'j'k'}d_{ijk}O_{i'i}O_{j'j}O_{k'k}\,.\label{Jd}
}
Thus, if one wants to implement an orthogonal transformation $O$ in the NHDM bilinear space, 
one has only the following two options:
\eq{
	\text{$O$ represents a }
	\left\{
	\begin{array}{cl}
		\text{family transformation if and only if}&\quad J=1 \text{ and } J_d=1\,,
		\cr
		\text{GCP if and only if}&\quad J=-1 \text{ and } J_d=1\,.
	\end{array}
	\right.
}
These criteria exclude, for instance, the case of $O=-\id$ mentioned above.

\section{Conclusions}

In order to exhaustively explore all phenomenological options offered by $N$ Higgs doublets,
one must perform an efficient scan in the immense parameter space, distinguishing
truly different models from mere basis changes.
The 2HDM and 3HDM experience shows that it can be efficiently done 
in the bilinear space.
In particular, it provides an intuitive geometric picture 
that generalized $CP$ transformations induce an odd number of mirror reflections
while the family symmetries always produce pure rotations.
The two types of transformations can be immediately distinguished by computing the determinant
and checking whether it is $-1$ or $+1$.

It was recently pointed out that this intuition fails already for 4HDM,
where all allowed transformations are represented in the bilinear space
with orthogonal matrices $O$ such that $\det O = +1$.
We cure this disturbing loss of intuition by replacing the determinant with the quantity $J$
given in Eq.~\eqref{O} and proving that $J= +1$ for Higgs family symmetries
and $J= -1$ for all GCPs.
Also, to remove the invalid transformations which can also lead to the same values of $J$, 
we also require that another quantity $J_d$ defined in \eqref{Jd} is equal to 1.
This opens up the possibility to imposes symmetries on NHDM scalar sector directly in the bilinear space:
only those orthogonal transformations $O$ which lead to $J_d = 1$ and $J = +1$ (basis changes)
or $J = -1$ (GCPs) can be used.
These results hold for any number of the Higgs doublets $N$.

\subsection*{Acknowledgments}
I.P.I.\ acknowledges funding from the Portuguese
\textit{Fun\-da\-\c{c}\~{a}o para a Ci\^{e}ncia e a Tecnologia} (FCT) through the FCT Investigator 
contract IF/00989/2014/CP1214/CT0004 under the IF2014 program,
and through the contracts UID/FIS/00777/2013, CERN/FIS-NUC/0010/2015, and PTDC/FIS-PAR/29436/2017,
which are partially funded through POCI, COMPETE, Quadro de Refer\^{e}ncia
Estrat\'{e}gica Nacional (QREN), and the European Union (EU).
I.P.I.\ also acknowledges support from the Government of the Russian Federation via Grant 2019-220-07-397
and from the National Science Center, Poland, via the project Harmonia (UMO-2015/18/M/ST2/00518).
C.C.N.\ acknowledges partial support by brazilian Fapesp, grant 2014/19164-6, and
CNPq, grant 308578/2016-3.

%%%%%%%%%%%%%%%%%%%%%%%%%%%%%%
\appendix

\section{The necessary and sufficient conditions on the orthogonal transformation}

Here we want to establish not only necessary but also sufficient conditions on an orthogonal transformation $O$
in the bilinear space of NHDM to represent a viable transformation back in the space of $N$ doublets, either a basis change or a GCP.

Let us define two fully antisymmetric tensors with indices in $\RR^{N^2-1}$:
\begin{equation}
F_{ijk} = \frac{f_{ijk}}{\sqrt{N(N^2-1)}}\,, \quad 
G_{ijk} = \frac{f_{i'j'k'}O_{i'i} O_{j'j} O_{k'k}}{\sqrt{N(N^2-1)}}\,,
\end{equation}
$O \in O(N^2-1)$. Let us also introduce the scalar product of such tensors:
$\langle F, G\rangle \equiv \sum_{ijk} F_{ijk} G_{ijk}$,
as well as the norm associated with this scalar product: $\| F \|^2 = \langle F, F\rangle \ge 0$.
Since the norm is a sum of squares of individual entries, it can be zero only if the tensor is zero.

By construction, $\|F\|^2=1$ due to \eqref{O-R} and $\langle F, G\rangle = J$ in \eqref{O}.
Since $O^T O = \id$, it also follows that $\| G \|^2 = \| F \|^2 = 1$.
We have seen in the main text that if $O=R$ is generated by a basis change in the space of doublets,
that is, $O$ is a member of $\adj(SU(N))$ (adjoint representation of $SU(N)$ in the $\RR^{N^2-1}$ space), then
\begin{equation}
G_{ijk} = F_{ijk} \quad \Rightarrow \quad J = 1\,.
\end{equation}
Similarly, we have shown that if $O=R\cdot C^{(s)}$ corresponds to a GCP transformation on the doublets, then 
\begin{equation}
G_{ijk} = -F_{ijk} \quad \Rightarrow \quad J = -1\,.
\end{equation}

Now we want to prove the converse: if $J=\aver{F,G} = 1$, then $O \in \adj(SU(N))$ 
and therefore it can be represented as a basis change in the space of doublets.
Analogously, if $J=-1$, then $O$ corresponds to a GCP transformation.
As we explained in the main text, just requiring $J = \pm 1$ is not sufficient, 
and we want to find an additional condition to be imposed to make the converse statement valid.

First, since
\begin{equation}
\| F\mp G \|^2 = \| F \|^2 + \| G \|^2 \mp 2 \langle F, G\rangle
\,,
\end{equation}
we conclude that $G = F$ if $J=1$ and that $G=-F$ if $J=-1$.
In the first case, the transformation $O$ leaves the $SU(N)$ structure constants invariant.
The same is true for the second case if we replace $O\to -O$ and observe that $G_{ijk}(-O)=-G_{ijk}(O)$,
where we made the dependence on the matrix $O$ explicit.
So if $J=1$ for a given $O$, then $J=-1$ for $-O$.
Note that if $N$ is odd, both $\pm O$ belongs to $SO(N^2-1)$ whereas if $N$ is even, only one of $\pm O$ belongs to $SO(N^2-1)$.

The invariance of the structure constants of $SU(N)$ for $J=1$ by the matrix $O$ can be written as
\eq{
\label{f.inv}
f_{i'j'k'}O_{i'i}O_{j'j}O_{k'k}=f_{ijk}\,.
}
For $J=-1$, the replacement $O\to -O$ is understood and from here on $O$ denotes a matrix 
that satisfies the invariance \eqref{f.inv}.
The transformations in \eqref{R} or \eqref{Cs}, which we can generically denote by $O$,
can be understood as a map $\psi$ on the Lie algebra $su(N)$ of $SU(N)$:
\eqali{
\psi\colon su(N)&\to su(N)
\cr
t^i&\mapsto O_{ji}t^j\,.
}
The invariance \eqref{f.inv} implies that 
\eq{
[\psi(t^i),\psi(t^j)]=if_{ijk}\psi(t^k)\,,
}
and $\psi$ represents an automorphism of $su(N)$.

Now, all automorphisms of $su(N)$ are classified (see e.g.\ Ref.\,\cite{Grimus:1995zi}).
They form a group denoted as $\aut(su(N))$. The invariant subgroup $\inn(su(N))$ consists of \emph{inner} automorphisms induced by conjugation of the group itself as
\eq{
\psi_S(t^i)=e^{iS}t_ie^{-iS}\,,
}
where $S\in su(N)$.
In other words,
\eq{
\psi_S(t^i)=R_{ji}(S)t_j\,,
}
where $R_{ji}(S)\in\adj(SU(N))$.
Automorphisms that are not inner are \emph{outer} and for $su(N)$, $N\ge 3$, they can all be 
characterized from
\eq{
\aut(su(N))/\inn(su(N))\simeq \mathbb{Z}_2\,.
}
A nontrivial representative $\psi^\Delta$ that generates $\mathbb{Z}_2$ can be chosen exactly as the transformation \eqref{Cs3}, generalized for any $N$, induced by the opposite of the canonical CP transformation:
\eq{
\psi^\Delta(t^i)=-(t^i)^T=O_{ji}t_j\,,\quad
O=-C^{(s)}\,.
}
So we are only left with two options for $O$:
\eqali{
\psi_O\in\inn(su(N))&:\quad O=R\,,
\cr
\psi_O\not\in\inn(su(N))&:\quad O=\big(-C^{(s)}\big)\cdot R\,,
}
where $R\in \adj(SU(N))$.
For both cases $J=1$, but only the first option corresponds to basis transformations on $\phi_a$.
For the second option, it implies that $-O$, the one which produces $J = -1$, is a viable option as it corresponds to a GCP.

Finally, to resolve the $\pm 1$ ambiguity,
we can use the transformations properties for $d_{ijk}$:
\eq{
d_{i'j'k'}O_{i'i}O_{j'j}O_{k'k}=d_{ijk}
\quad
\text{ for both $O=R$ or $O=+C^{(s)}$.}\label{dOOOd}
}
To show the case for $C^{(s)}$, just use the definition of $d_{ijk}$ based on $t^i$ and of the automorphism $\psi^\Delta$.
We then define $J_d$ in \eqref{Jd} and require it to be $+1$ for the valid orthogonal transformations.

We also remark that the fact that valid GCPs must leave $d_{ijk}$ invariant as in \eqref{dOOOd} was also stated and proved
in \cite{Maniatis:2019yqb}. The proof was given only for 3HDM and with the aid of computer algebra program 
by considering special cases for $r_i$.
In our procedure, it is required only at the last step, when the structure of $O$ is already established up to the $\pm \id$ ambiguity and only the property of the standard CP transformation is necessary.
%We also note that the criterion $J_d = 1$ was not mentioned in \cite{Maniatis:2019yqb}.

\end{document}